\documentclass[preprint]{aastex}
\usepackage{spr-astr-addons}
\usepackage{url}
\usepackage{rotating}

\RequirePackage{color}

\newcommand{\emaila}{riccardo.campana@inaf.it}

\begin{document}

\title{New blazar candidates from the 9Y-MST catalogue detected at energies higher than 10 GeV} 

\shorttitle{New Blazar candidates from the 9Y-MST catalogue}
\shortauthors{R. Campana, E. Massaro}  

\author{R.~Campana\altaffilmark{}}
\affil{INAF/OAS-Bologna, via Gobetti 93, I-40129, Bologna, Italy. \\ \emaila} 
\and 
\author{E.~Massaro}
\affil{INAF/IAPS, via del Fosso del Cavaliere 100, I-00133, Roma, Italy}
\affil{In Unam Sapientiam, Roma, Italy}
\email{riccardo.campana@inaf.it} 

\begin{abstract}
We present a list of 24 new blazar candidates selected in a search for possible counterparts
of  spatial clusters of $\gamma$-ray photons in the recent 9Y-MST catalogue, at energies higher than 10~GeV and 
at Galactic latitudes higher than 20\degr. 
13 of these clusters are also included the preliminary release of the 4FGL catalogue of 
$\gamma$-ray sources. 
The search for possible counterparts is based on the possible associations of the clusters with
radio sources within a circle having a radius of 6\arcmin.
We then investigated the possible optical or mid-IR associations of these sources, checking 
if they show some properties typical of new blazar candidates.
\end{abstract}


\section{Introduction}\label{s:introduction} 

The use of spatial clustering algorithms in the $\gamma$-ray sky allows to extract
photon concentrations out of the diffuse background, which can be then associated
with high energy cosmic sources.
The multiwavelength analysis of the fields surrounding these {photon} clusters is useful to identify 
new sources that were missed in other surveys based on different selection criteria.

In a previous paper, we used the Minimum Spanning Tree (hereafter MST, \citealt{campana08,campana13}) 
method to produce a new catalogue of 1342 {photon} clusters, at Galactic latitudes 
$|b| > 20 \degr$ in the 
\emph{Fermi}-Large Area Telescope (LAT) dataset above 10 GeV and covering a 
9 years time interval from the beginning of the mission  (9Y-MST, \citealt{campana18}).
The large majority of them correspond to sources already detected in previous 
searches, and generally were found to be closely associated with known BL Lac 
objects, Flat Spectrum Radio Quasars (FSRQ) and other Active Galactic Nuclei (AGN).

The 9Y-MST includes also 249 clusters without any correspondence with previously known
high energy sources.
As discussed in \citet{campana18} it is possible that a fraction of these unassociated 
{photon} clusters may be \emph{spurious}, i.e. random fluctuations in the photon density,
not related to any physical counterpart.
However, many of these clusters were found having characteristic parameters quite close to those of the 
associated ones and therefore they may be actually related to cosmic sources not yet studied and classified.
We searched for possible counterparts and obtained a sample of 24 {spatial clusters of photons} whose centroids
have an angular separation from radio sources comparable to those found for 
clusters already associated with optical/IR objects exhibiting properties typical of
blazars.
In a series of papers \citep{bernieri13,paperI,paperIV,paperIII,paperII,campana17} we 
described the results of similar searches of possible new blazar counterparts of $\gamma$-ray
clusters found by means of MST, which were, with a very few exceptions, confirmed by 
subsequent analyses.

In February 2019 the Fermi-LAT collaboration released a preliminary version 
of the 4FGL catalogue \citep{4FGL}, including 5098 $\gamma$-ray sources and reporting confirmed
and candidate counterparts.
{In our selected sample of photon clusters, there are 13 out of 24 that correspond to 4FGL sources. 
Although they can be considered as \emph{bona fide} $\gamma$-ray sources, we mantained them in the sample to allow for a comparison of 
their observational properties with respect to other candidate sources.}

The outline of this paper is as follows. 
In Section~\ref{s:criteria} and \ref{s:wise} we describe the criteria adopted for 
selecting the counterparts and present the resulting sample of new blazar candidates, 
in Section~\ref{s:ctrp} the main properties of individual sources are given and in 
Section~\ref{s:conclusions} the results are summarized and discussed.

\section{Cluster parameters and counterpart selection criteria}
\label{s:criteria}

For the preparation of the 9Y-MST catalogue we considered LAT data (Pass 8R2) above 10~GeV, 
covering the whole sky in the 9 years time range from the start of mission (2008 August 04) 
up to 2017 August 04, were downloaded from the FSSC 
archive\footnote{\url{http://fermi.gsfc.nasa.gov/ssc/data/access/}}.
Standard cuts on the zenith angle, data quality and good time intervals were applied.

MST application and cluster selection criteria are described in the 9Y-MST paper \citep{campana18}, and
therefore we describe here only the cluster parameters useful for source association
and the search of possible counterparts.
{MST starts from considering the photon arrival directions in a given field as points (\emph{nodes}) in a 2-D reference frame, and constructs a particular graph, the \emph{minimum spanning tree}, connecting them with weighted \emph{edges}. The edge weight is the angular distance between a pair of photon. Then, all the edges with a length below a threshold $\Lambda_\mathrm{cut}$ (expressed as a fraction of $\Lambda_m$, i.e. the average of all the edge lengths in the whole graph) are removed, as well as the remaining sub-trees with a number of nodes below a threshold $N_\mathrm{cut}$. We refer to \cite{campana08,campana13} for details.}

{The main parameters of a cluster are its \emph{photon number} $n$ and the \emph{clustering
factor} $g$. 
The latter is defined as the ratio between the mean photon distance in the whole field $\Lambda_m$ to the mean distance in the cluster itself ($\Lambda_{m,k}$, i.e. the average of all the edge lengths in the specific $k$-th sub-tree).
This is a measure of the ``clumpiness'' of the cluster in consideration.}
The derived parameter $M$, the so called \emph{magnitude} \citep{campana13}, is defined as the
product $M = n g$ and was found to be related to the statistical significance
of clusters.

Another parameter is the \emph{cluster centroid} whose coordinates are computed as a
weighted mean of the coordinates of the photons in the cluster.
According to \citet{campana13}, applying as a weight the inverse of the square of the distance 
to the closest photon, it results a better agreement with the positions given by maximum likelihood {\citep[ML,][]{mattox96}} analysis.
However, in the case of clusters with a low number of photons the weighted centroid
may be biased if in the cluster there is a pair of photons much closer than the others:
the centroid will be then located very close to this pair and the use of the unweighted
mean can be more reliable.

Two other useful parameters are the \emph{maximum radius} $R_\mathrm{max}$, defined 
as the angular distance between the centroid and the farthest photon, that gives information 
on the overall extension of the cluster, and the \emph{mean radius} $R_m$, the radius of the 
circle centred at the centroid and containing 50\% of photons in the cluster, that 
for a point-like source should be smaller than or comparable to the 68\% containment 
radius of instrumental Point Spread Function \citep[PSF, see][]{ackermann13}. 

The search for possible counterparts of unassociated clusters was aimed to select a sample
of objects exhibiting one or more interesting properties to be considered blazar candidates.
On the basis of a positional matching between the 9Y-MST and the 3FHL \citep{ajello17}
catalogues, \citet{campana18}
found that more than 99\% of associations are within an angular separation $\delta <$ 6\arcmin, computed
using the cluster centroid and the 3FHL coordinates, and this figure
was used for searching other associations with lists of known blazars or candidates.
In the present work we adopted the same criterion and defined a search region having
a radius of 6\arcmin\ within to select new candidates.

Our first step was to extract from the NVSS\footnote{\url{https://www.cv.nrao.edu/nvss/}} (1.4 GHz) and 
SUMSS21\footnote{\url{http://www.astrop.physics.usyd.edu.au/sumsscat/}} (0.835 GHz) catalogues --- which together cover the entire sky ---
all the radio sources found in the searching region, to investigate if at least one of them would exhibit typical 
blazar features.
For the association we used also the more severe positional criterion $\delta < R_\mathrm{max}$. 
However, in the case of low $g$ clusters this choice could be problematic because 
$R_\mathrm{max}$ may have large variations depending on the selection parameters.
A more detailed analysis of the cluster structure was therefore performed to obtain a 
good association.
We also extended the search for data to optical and X-ray bands in order to 
verify whether radio counterparts might be associated with sources exhibiting
some blazar properties.
We thus obtained a sample of 24 $\gamma$-ray clusters that is listed in Table 
\ref{t:sources} together with the coordinates and names of selected candidate counterparts.
In the next Section \ref{s:ctrp} the properties of these candidates are presented individually.

\section{\emph{WISE} colours}\label{s:wise}

{
After the selection of radio candidates, we searched for possible optical and mid-IR \citep[AllWISE catalogue,][]{cutri13}  
counterparts having coordinates within their uncertainty radii, typically 
few arcseconds, and tested if their photometric data are compatible with a blazar nature. 
}
More specifically, a very useful test is to check if the mid-IR colours are located 
within the \emph{WISE Gamma-ray Strip} according to the definition of \cite{massarodabrusco16}.

\begin{figure}[h]
\centering 
\includegraphics[width=0.48\textwidth]{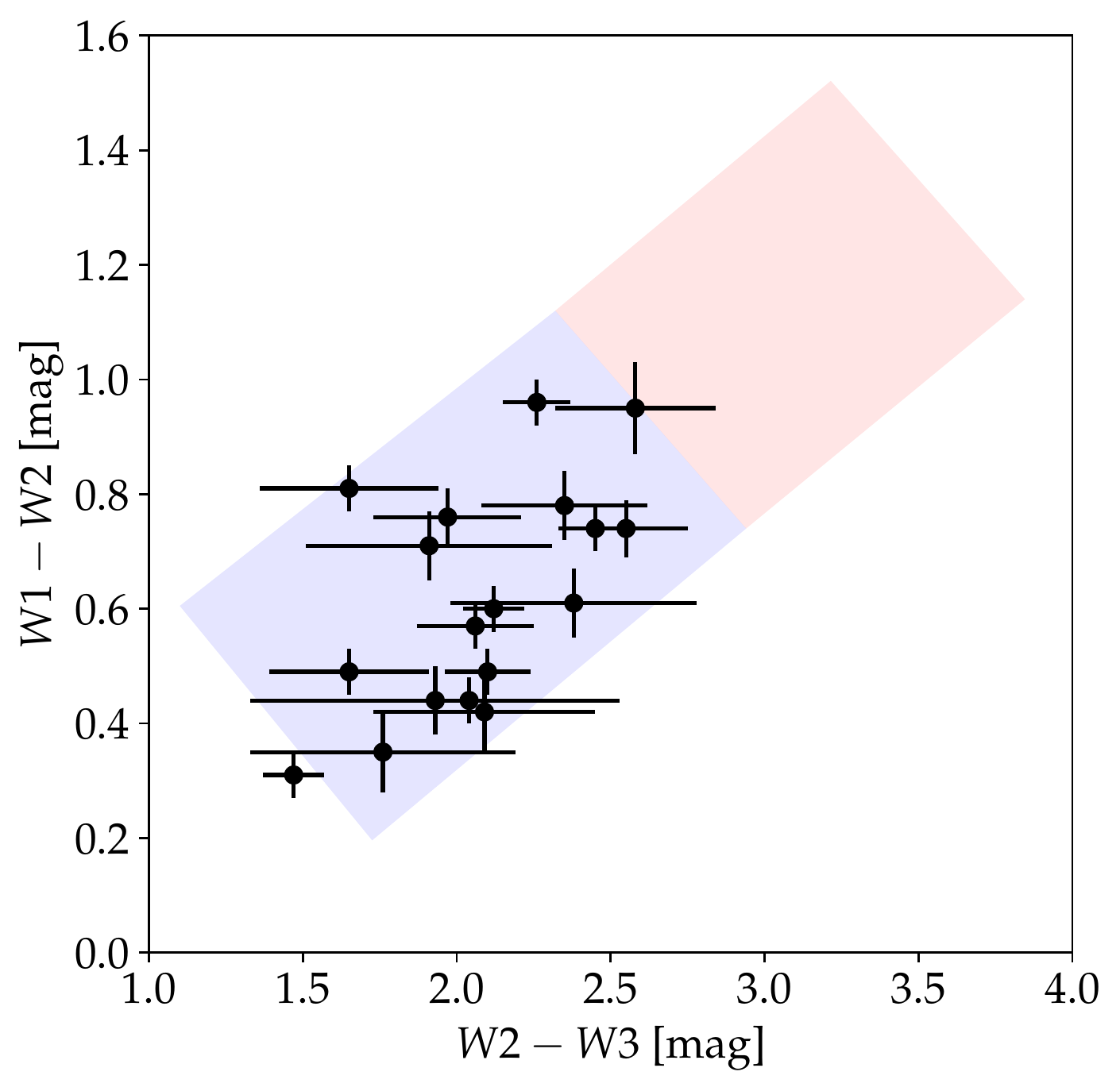}
\caption{Plot of the infrared colours of 18 new candidate blazars with data from the 
AllWISE catalogue.
The shaded regions define the WISE Blazar locus \citep{massarodabrusco16} in which  $\gamma$-ray 
loud blazars are present. 
The blue-shaded region represents the locus where there is a concentration of BL 
Lac objects, while the red-shaded region correspond to FSRQ objects.}
\label{f1}
\end{figure}

We obtained three-band infrared photometric data from the AllWISE catalogue: 
for 18 out of 24 cluster candidate counterparts in the three
bandpasses $W1$ [3.4 $\mu$m], $W2$ [4.6 $\mu$m] and $W3$  [12 $\mu$m], while 
in the lowest frequency band $W4$ [22 $\mu$m] photometric data is available 
only for five of these sources.
The remaining 6 sources were detected only in the $W1$ and $W2$ bands and therefore
only one colour is available.  
{Note also that the uncertainty of the $W3$ magnitude of SDSS J174402.91$+$463740.7 
is not given in the VIZIER database, and therefore the reported magnitude should be considered as an upper limit.}
We computed the $W1 - W2$ and $W2 - W3$ colours without reddening correction,
since all the considered sources have a Galactic latitude higher than 20\degr.
In any case, its largest effect on the colours is of only a few hundredths of
magnitude, quite lower than typical uncertainties. 
The resulting two WISE colour plot that is shown in Figure~\ref{f1}.
In this plane $\gamma$-ray blazars are essentially concentrated within the 
two coloured areas, defined in the figures reported in \citet{massaro13}, 
\citet{dabrusco14}, and \citet{massarodabrusco16}:
BL Lac objects are mainly concentrated in the blue area, while FSRQ are mostly 
found in the red one.
In their data there is no definite boundary between BL Lac and FSRQ regions, and 
the given separation is only indicative.
All our candidates have colours matching very well the BL Lac region.
Note, in particular, the very close similarity between our plot and the one given in
Figure 2 of \citet{massaro13}.

Recently, two new catalogues of blazar candidates has been published \citep{dabrusco19}
based on \emph{WISE} infrared data. WIBRaLS2 contains sources with 4-band photometric data, spatially matched
to radio-loud sources, while KDEBLLACS collects radio-loud sources with 3-band \emph{WISE} data only, 
with mid-infrared colors similar to $\gamma$-ray confirmed BL Lacs.
Association of some of the clusters on our selected samples to those two catalogues are discussed in the following Section.

\section{Properties of counterparts to individual clusters}
\label{s:ctrp}

\subsection{9Y-MST J0013$-$3222}  
There are three NVSS radio sources, two of them are also in SUMSS21, within 6\arcmin.
The brightest, at the closest angular distance to the $\gamma$-ray postion 
equal to 2\farcm71, is NVSS J001339$-$322445.
It is associated with the X-ray counterpart 1RXS J001338.8$-$322442.

\begin{figure}[htb]
\centering 
\includegraphics[width=0.48\textwidth]{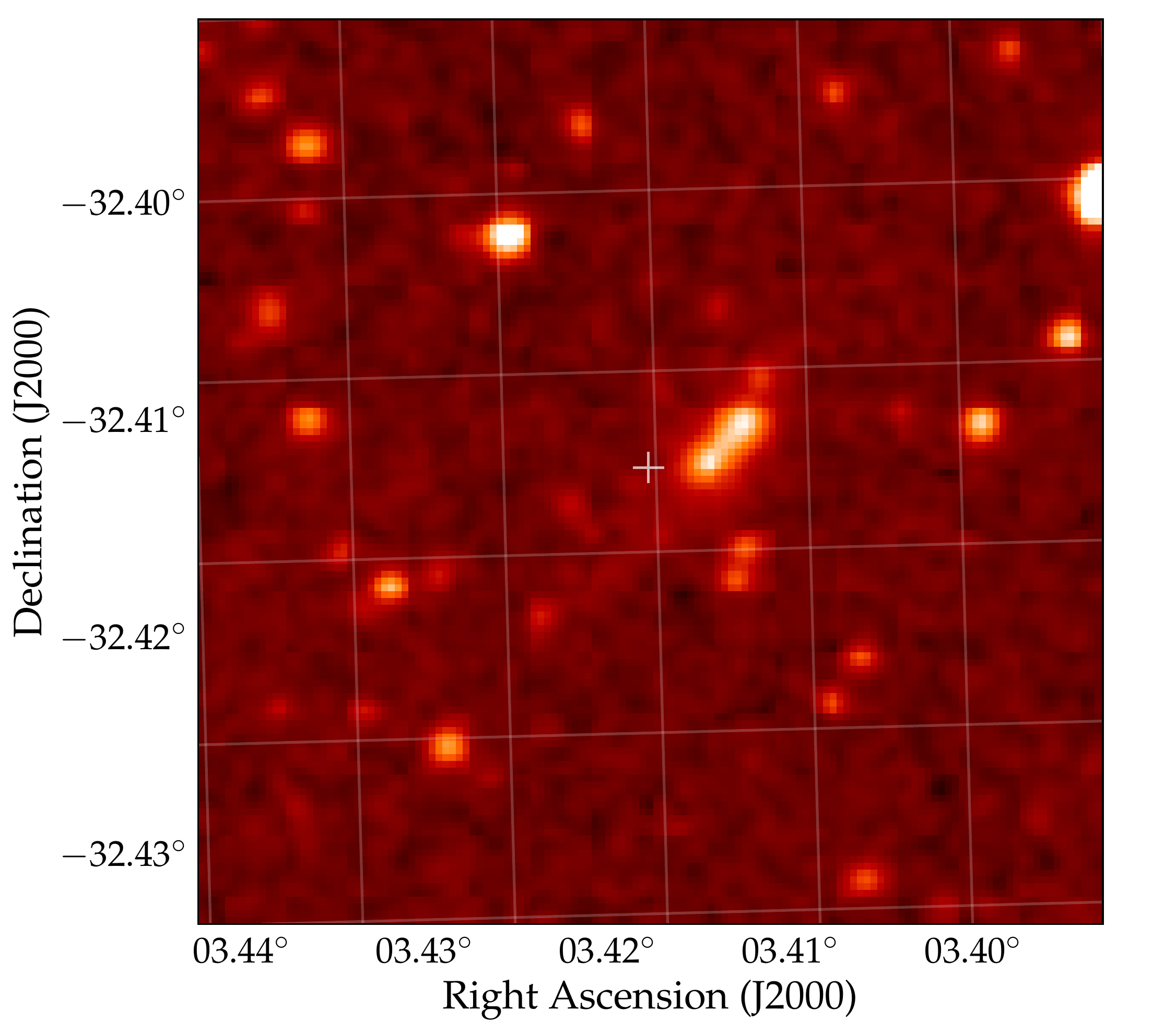}
\caption{ESO DSS $R$-band image of the field surrounding NVSS J001339$-$322445 (white cross).}
\label{fig_A1}
\end{figure}

The ESO DSS\footnote{\url{http://archive.eso.org/dss/dss}} $R$-band image (Figure~\ref{fig_A1})
is rather peculiar, showing two knots, or galactic nuclei,
about 12\arcsec\ apart, of similar brightness embedded in an elongated nebular structure. 
A fainter weaker knot is in the middle the main two.
The nebular structure has an extended straight feature, aligned in the same direction
of the line connecting the knots, resembling a faint jet.  

2dF\footnote{\url{http://www.2dfgrs.net}} spectra of both knots are available: 
the spectrum of quality $Q = 4$ is typical of an elliptical galaxy without emission 
lines and a rather low Ca H+K break ratio;
the reported redshift is $z = 0.2598$. 
The radio spectrum is clearly steep at low frequencies: the spectral index from
SUMSS21 (0.843 GHz) and NVSS (1.4 GHz) is equal to $-1.2$, rather close to the value 
of $-1.0$ reported at frequencies between 72 and 231 MHz in the GLEAM EGC catalogue
\citep{hurley-walker17}
However, there is an indication of a flattening at high frequencies since between
1.4 and 4.85 GHz the spectral index is $-0.34$.

This source is also present in the ROXA sample \citep{turriziani07} 
and its classification is uncertain Radio Galaxy or BL Lac object.

\subsection{9Y-MST J0024$+$2401}  
This cluster has a well established correspondence at a distance of 0\farcm8 with 
the $\gamma$-ray source 4FGL J0024.1$+$2402,  
for which no possible counterpart is reported.

\begin{figure}[htb]
\centering 
\includegraphics[width=0.48\textwidth]{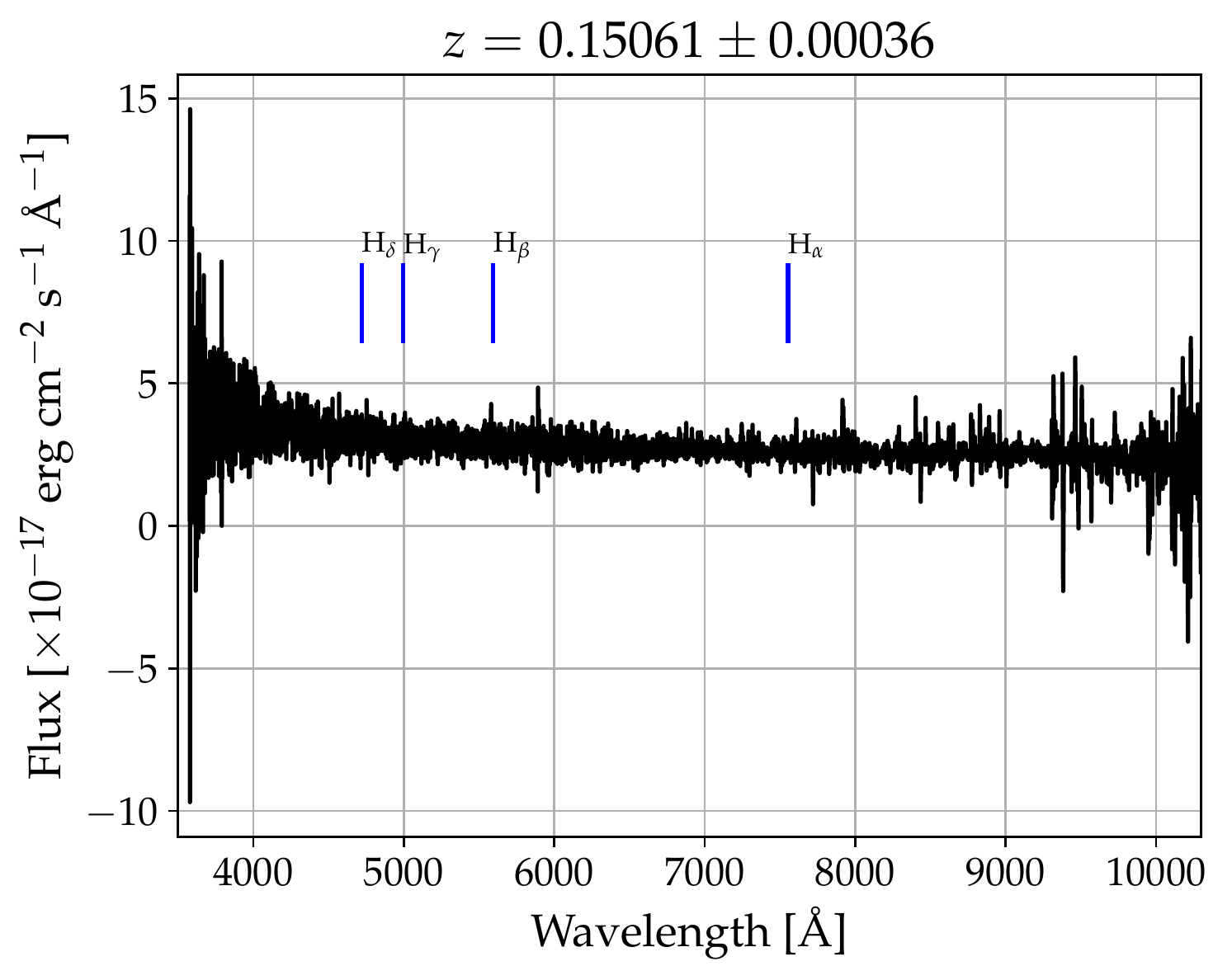}
\caption{Sloan Digital Sky Survey optical spectrum (plate no. 7661, fiber no. 398) of 
SDSS J002406.10+240438.3.
Vertical blue segments mark the position of possible Balmer emission lines found
by the SDSS automatic analysis.}
\label{fig_B1}
\end{figure}

There are a few nearby NVSS sources, but the closest and brightest is at 2\farcm97
and has the interesting optical counterpart SDSS J002406.10+240438.3:
its spectrum (Figure~\ref{fig_B1}) has some features detected by the automatic SDSS
procedure with an estimated $z = 0.151$, however with small significance (low $\Delta \chi^2$).
It appears as a featureless flat continuum with a blue excess with respect to
the typical spectrum of an elliptical galaxy.
Furthermore, it corresponds to the mid-IR source AllWISE J002406.10+240438.6, having
the colours $W1 - W2 = 0.87$ and $W2 - W3 = 2.22$, close to the centre of the 
\emph{WISE} Gamma-ray Strip \citep{massarodabrusco16} 
{and it is included in the KDEBLLACS sample \citep{dabrusco19}.}

The association with a new BL Lac object can be considered as a robust result.

\subsection{9Y-MST J0122$+$1033}   
There is only one NVSS source within a distance of 6\arcmin, that has a very likely optical 
counterpart in SDSS, a starlike source, but unfortunately no spectrum is available.
The colour $u - r = 0.63$ is, however, rather blue and well compatible with a quasar.
\cite{richards15} and \cite{brescia15} included this object in their catalogues
of candidate quasars, the former also based on the mid-IR $WISE$ data \citep{cutri13}.

It is therefore a likely blazar candidate, but more information is necessary to confirm 
the nature of this object.

\subsection{9Y-MST J0127$+$1737}  
This is a ``poor'' cluster, having only 4 photons, but with a quite high $g$ that implies 
a significant $M$ value.
There is only one rather faint NVSS radio source within a distance of 6\arcmin.
At about 5\arcsec, compatible with the NVSS positional uncertainty, there is a possible
faint counterpart in SDSS with $u - r = 0.64$, but no spectrum is available.
This source has an interesting mid-IR counterpart in the $WISE$ sky with the colours
$W1 - W2 = 0.78$ and $W2 - W3 = 2.35$ placing it in the middle of the Gamma-ray blazar strip 
\citep{massarodabrusco16} in the BL Lac object section {and and satisfying the criteria for inclusion
in the KDEBLLACS sample \citep{dabrusco19}.}

It is reported in the quasar candidate lists by \cite{brescia15} and \cite{richards15}, 
who give a photometric redshift estimate around 0.5.

It is therefore a good High Energy Peaked BL Lac (HBL) candidate at high redshift which 
must be confirmed by spectral data.

\subsection{9Y-MST J0143$-$0122} 
\begin{figure}[htb]
\centering 
\includegraphics[width=0.48\textwidth]{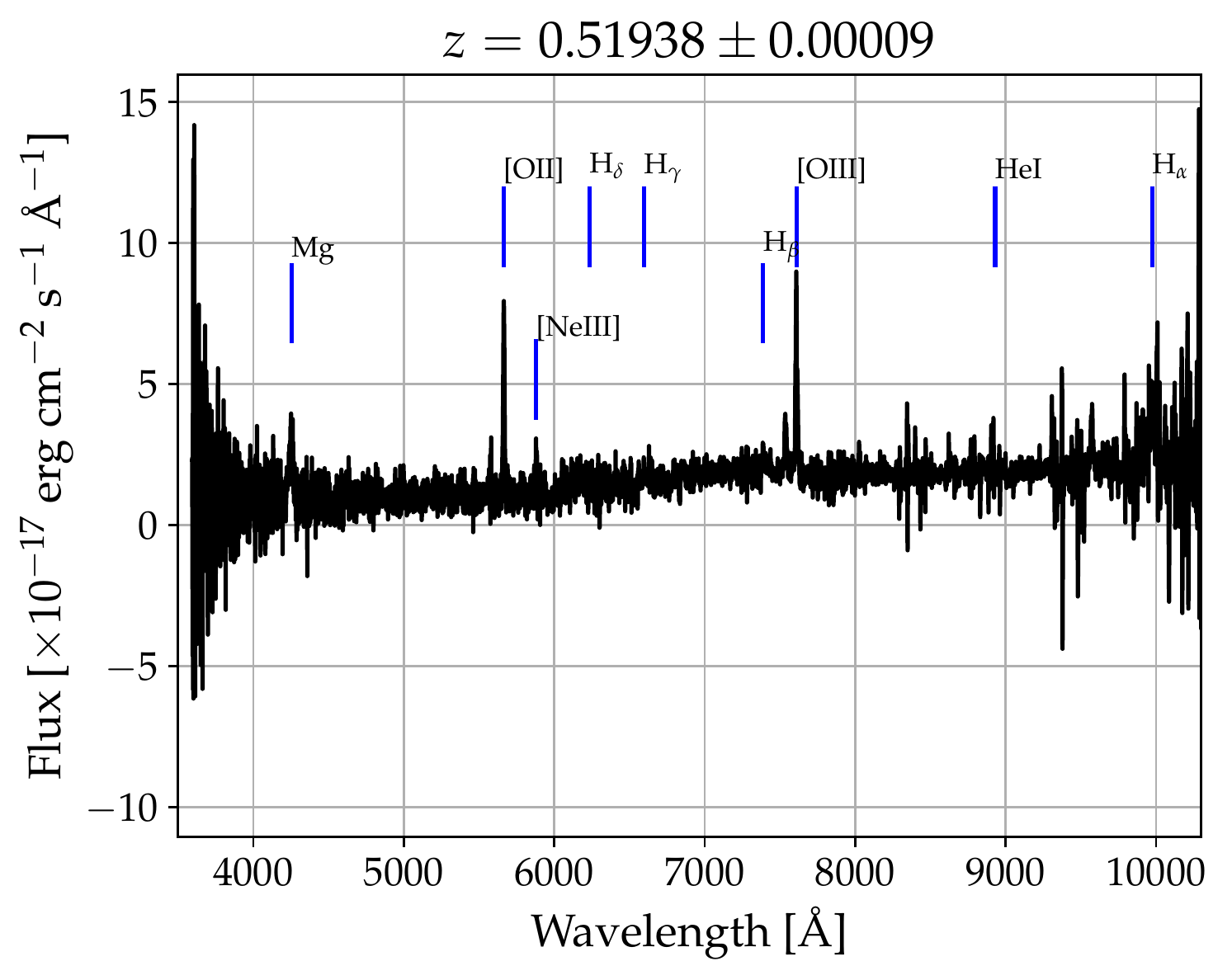}
\caption{Sloan Digital Sky Survey optical spectrum (plate no. 4350, fiber no. 783) of 
the optical counterpart associated to NVSS J014317$-$011858.
Vertical blue lines mark some of the emission features automatically detected in SDSS.
}
\label{fig_C1}
\end{figure}

There are several radio sources within a distance of 6\arcmin\ and therefore the possibility
of a multiple association cannot be excluded.
The brightest radio source is NVSS J014317$-$011858, also reported as PKS J0143$-$0119 and 
4C~$-$01.09. 
It appears in NVSS but results fragmented in several components of different brightness
in FIRST. 
VLA images \citep{reid99,roberts15} have a core-jet structure
with some central bright knots and the peak at extreme west was associated
with the optical counterpart by \cite{lacy00}, a faint object at $z = 0.5194$, confirmed
by a SDSS spectrum and exhibiting several broad emission lines (Figure~\ref{fig_C1}).
As noticed by \cite{roberts15}, the nature of this object is still unclear.
In particular, it is not established whether it has or not a flat spectrum component
embedded in a steep spectrum extended emission.
An indication for this possibility is provided by the NRAO 
VLBA\footnote{\url{http://www.vlba.nrao.edu/cgi-bin/vlba_calib.cgi}} 
calibration data and images at 2.3 and 8.6 GHz having only a single compact component 
with a possibly inverted spectrum.
The search in the AllWISE catalogue gives a poorly resolved source J014316.73$-$011900.6 
with mid-IR colours $W1 - W2 = 0.61$ and $W2 - W3 = 2.38$ 
well located in the BL Lac section of the  \emph{WISE} Gamma-ray Strip \citep{massarodabrusco16}.

For the sake of completeness we mention the presence at 6\farcm6 of another blazar candidate, 
the source 2WHSP J014347.1$-$01260, located just outside the $R_\mathrm{max}$ circle, which has an 
AllWISE counterpart with only the $W1 - W2 = 0.18$ that places it in a marginal position of 
the WISE strip.

\subsection{9Y-MST J0202$+$2942} 
This cluster corresponds to 4FGL J0202.4$+$2943, but no counterpart is given.
There is only one NVSS radio source in its environment which has a clear optical starlike 
counterpart.
Unfortunately, no spectrum is available in SDSS but DR15 photometric data give $u - r = 0.86$.
It was reported in the quasar samples by \cite{brescia15} and \cite{richards15}.
The AllWISE counterpart has the colours $W1 - W2 = 0.73$, $W2 - W3 = 2.44$, $W3 - W4 = 2.04$
which locate it well within the BL Lac segment of the WISE blazar strip used for selecting 
the WIBRaLS and WIBRaLS2 samples \citep{dabrusco14,dabrusco19}.

It is therefore likely that this source is a good HBL candidate, whose proper classification
requires an optical spectrum.

\subsection{9Y-MST J0332$+$8227}  
This cluster corresponds to  4FGL J0333.1$+$8227 and was related to 1RXS J033208.6$+$822654,
classified as `bcu' (i.e. blazar candidate of uncertain type).
There is only one NVSS radio source within 6\arcmin\ that is at a very close angular distance.
The corresponding optical source is very faint and there is no spectral information; however
there is a mid-IR counterpart with the colours $W1 - W2 = 0.81$, $W2 - W3 = 1.64$, 
$W3 - W4 = 3.15$, that place this object in a rather marginal position with respect
to the WISE blazar strip, but compatible when the errors are taken into account.

\subsection{9Y-MST J0557$+$7705}  
This cluster corresponds to the preliminary source FL8Y J0557$+$7705 
but it is not included in the final 4FGL catalogue. 
There are two NVSS radio sources within 6\arcmin: one is at the close angular distance of
0\farcm9, while the other is at 5\farcm9, close to $R_\mathrm{max}$ but quite higher than $R_m$.
We thus selected the former one as the most likely counterpart.
There is no clear source in POSS, but the GAIA DR2\footnote{\url{https://gea.esac.esa.int/archive/}} reports an object
at about 1\arcsec\ having a Gmag of 19.2.
There is a much brighter counterpart in the mid-IR, AllWISE J055721.46+770443.2, detected
in three of the four bands and having the colours  $W1 - W2 = 0.96$, $W2 - W3 = 2.26$,
which locate it close the centre of the \emph{WISE} Gamma-ray Strip in the transition
region between BL Lac objects and FSRQ, {but it is also included in the KDEBLLACS sample \citep{dabrusco19}.}

This source was included in the AGN catalogue based on mid-IR data by \cite{secrest15}.

The available data do not allow a safe classification of the proposed counterpart and more
information is necessary to confirm its blazar nature.

\subsection{9Y-MST J0650$-$5146} 
\begin{figure}[htb]
\centering 
\includegraphics[width=0.48\textwidth]{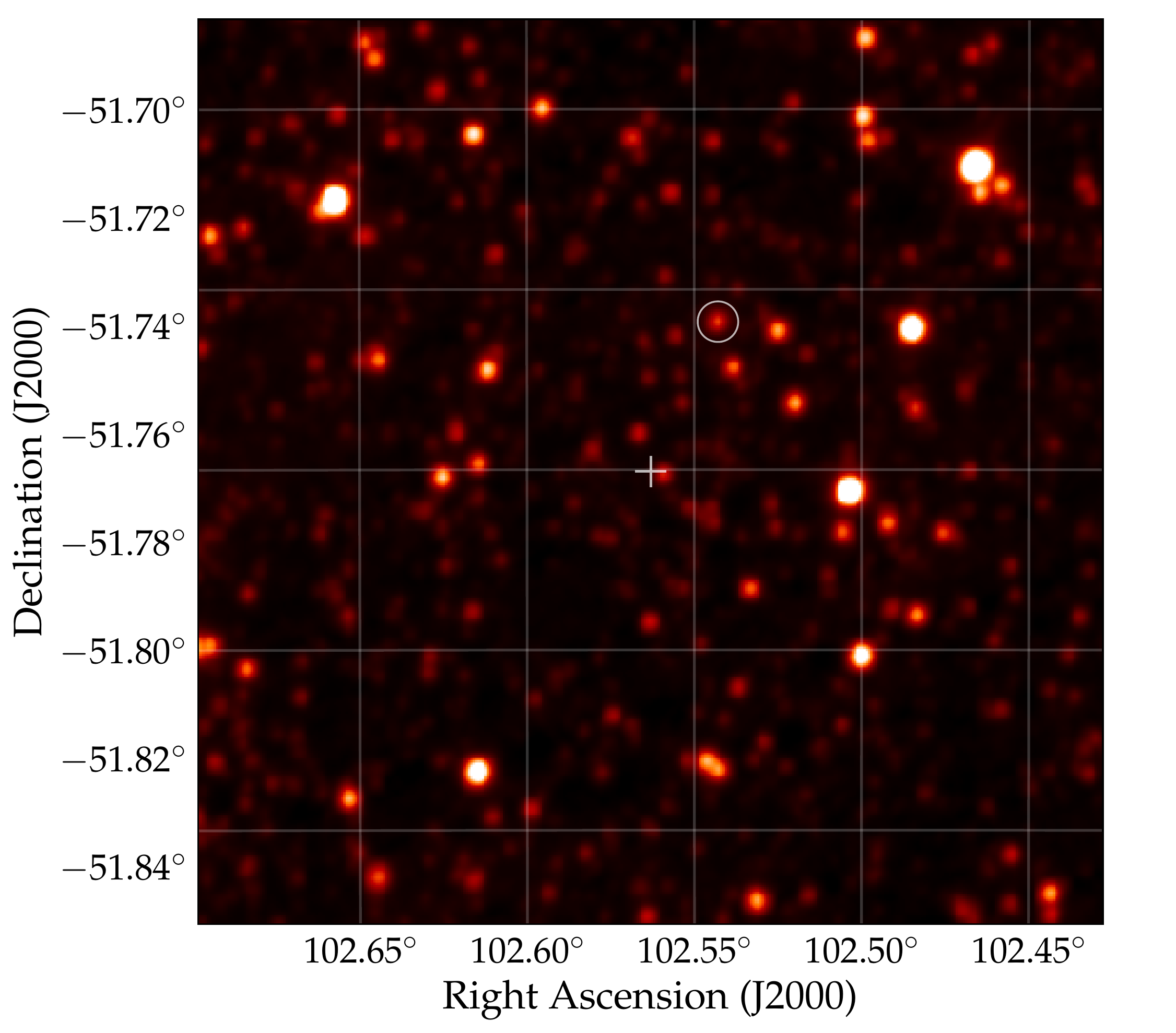}
\caption{WISE band 1 (3.4 $\mu$m) image of the field surrounding 9Y-MST J0650$-$5146 (whose centroid 
is marked by the cross). The circle marks its most likely counterpart, as discussed in the main text.}
\label{fig_D1}
\end{figure}

This cluster corresponds to 4FGL J0650.2$-$5144, but no counterpart is reported.
It lies at a low Galactic latitude in a rather crowded field.
There are two SUMSS21 radio sources within 6\arcmin: the brighter one is at 1\farcm8 and 
the other at 4\farcm5.
The former has a reasonable optical counterpart, while the latter does not correspond to
any bright enough source in POSS.
No spectral data are available for this source, but it has a well established counterpart
in the mid-IR with 3 band photometric data  (Figure~\ref{fig_D1}).
The resulting colours ($W1 - W2 = 0.61$, $W2 - W3 = 2.11$) place it among the BL Lac segment
of the \emph{WISE} Gamma-ray Strip {and it is included in the KDEBLLACS catalogue.}

We notice also that there is a 3FGL source at about 32\arcmin, without a correspondent cluster
in the 9Y-MST catalogue, but that is likely associated with the blazar candidate 2WHSPJ064710.0-51354.
Moreover, this latter source corresponds to 4FGL J0647.0$-$5138, different from the one associated with
our cluster.

\subsection{9Y-MST J0752$+$7120}  
Several NVSS sources are in the surroundings of this cluster.
Considering that it is the most compact one in the sample with $R_\mathrm{max} =$ 2\farcm9 and,
therefore, if we limit the search radius to this value, the number of radio sources is reduced 
to 2, of which only one has an interesting optical counterpart.
This source has a good positional correspondence with the X-ray source 1RXS J075225.0$+$712048
classified a quasar by \cite{flesch16}.
The only available WISE colour $W1 - W2 = 0.5$ is compatible with the BL Lac portion
of the blazar strip \cite{massarodabrusco16}.

It appears an interesting blazar candidate, but more data are necessary for a correct classification.

\subsection{9Y-MST J0947$+$1120}   
\begin{figure}[htb]
\centering 
\includegraphics[width=0.48\textwidth]{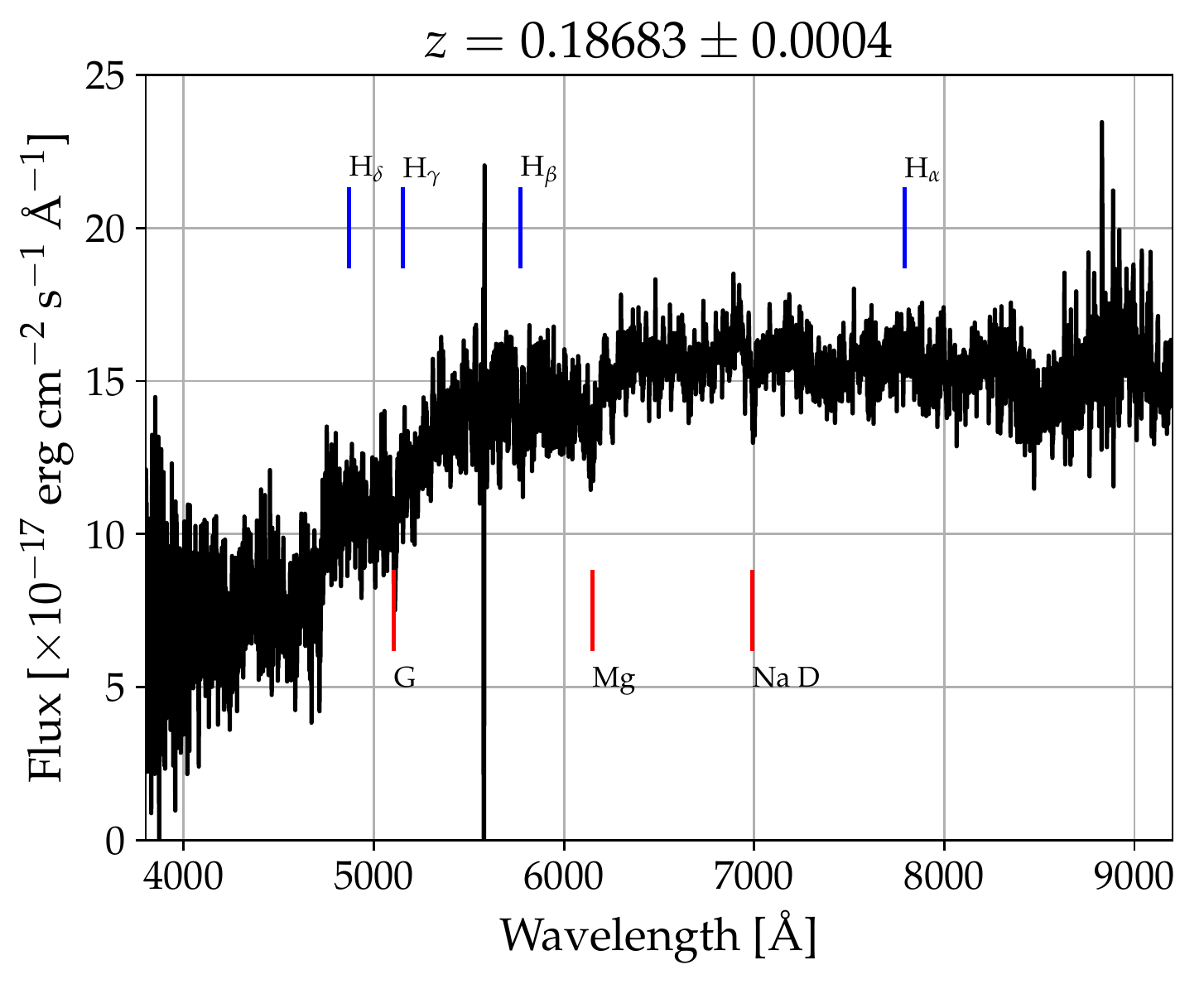}
\caption{Sloan Digital Sky Survey optical spectrum (plate no. 1742, fiber no. 37) of
SDSS J094745.91$+$112021.9.
Vertcal red lines indicate some of the absorption features reported in SDSS. 
No significant emission line is apparent. Balmer series location is marked by blue vertical lines.
}
\label{fig_E1}
\end{figure}

There are four NVSS sources with fluxes ranging from 3 to 48 mJy, and all are within the
$R_\mathrm{max} = 4\farcm2$ circle. 
The brightest source is the elliptical galaxy SDSS J094745.91$+$112021.9 at $z = 0.187$, and
its spectrum (Figure~\ref{fig_E1}) presents some clear absorption lines; this object was 
already included in the sample of BL Lac candidates by \cite{plotkin08}.
It is also associated with the X-ray source 2RXS J094746.1+112030. 
Unfortunately, a close very bright star makes difficult a good mid-IR photometry,
thus there is no information on its position in the \emph{WISE} colour plot useful for
the blazar classification.

None of the other radio sources has optical counterparts useful for unraveling
their nature and thus they are not further considered in the present work.

\subsection{9Y-MST J1003$-$2139}  
\begin{figure}[htb]
\centering 
\includegraphics[width=0.48\textwidth]{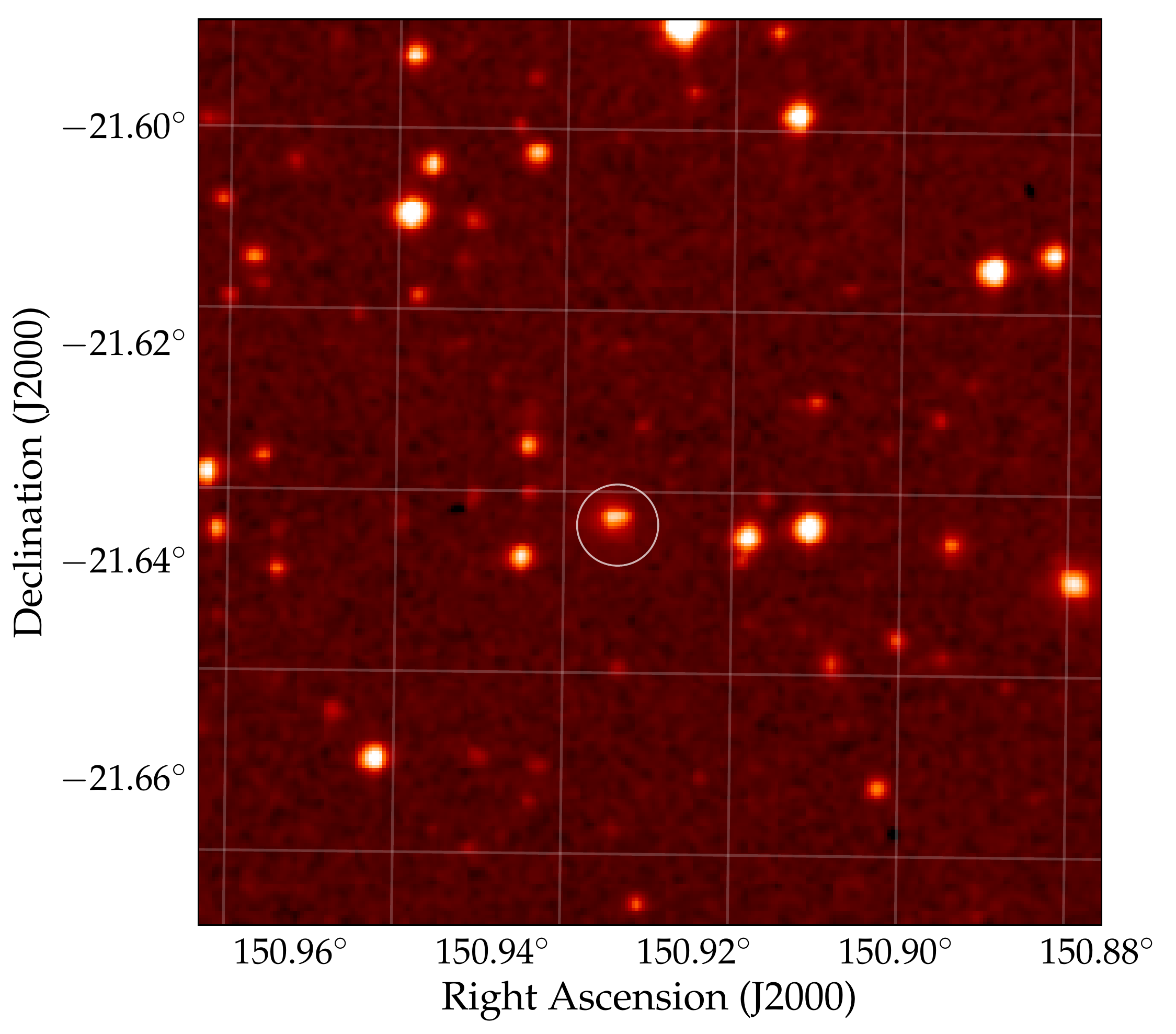}
\caption{ESO DSS $R$-band image of the field surrounding 1RXS J100342.0$-$213752. 
The peculiar optical counterpart discussed in the text is marked by the white circle.}
\label{fig_F1}
\end{figure}

This cluster corresponds to 4FGL J1003.6$-$2137 which is associated with the `bcu' source 
1RXS J100342.0-213752.
There is only one NVSS source close to the cluster centroid corresponding to this RASS 
counterpart, also detected by XMM.
The associated optical object is peculiar because of its elongated shape, unresolved
in the available images (Figure~\ref{fig_F1}); it is classified as `extended' in the HYPERLEDA 
database \citep{makarov14}.
It is possible that it is a very close pair of starlike objects and for this reason 
its optical magnitude might be brighter than the real value.
It has a relatively bright AllWISE counterpart having colours $W1 - W2 = 0.31$, 
$W2 - W3 = 1.47$, that place this object at the lower end of the \emph{WISE} BL Lac strip.

\subsection{9Y-MST J1636$-$0456} 
This cluster corresponds to 4FGL J1636.5$-$0454, which is associated to a NVSS source
classified as `bcu'.
There are two NVSS sources close to the cluster centroid, but only one is at an angular
distance lower than $R_\mathrm{max}$, the same reported in the 4FGL catalogue, which has also the optical 
counterpart SDSS J163632.084$-$045506.0, a galaxy without available spectral data.
The $u - r$ colour is equal to 2.41, large for a blazar-like object according to \cite{massaro12}.
It has a mid-IR counterpart whose colours ($W1 - W2 = 0.56$, $W2 - W3 = 2.06$) place 
it well among BL Lac objects in the  \emph{WISE} Gamma-ray Strip {and in the KDEBLLACS sample.}

\subsection{9Y-MST J1646$-$0942}  
This cluster corresponds to 4FGL J1646.0$-$0942 and the RASS source 1RXS J164602.3$-$094113
is indicated as counterpart.
There are three possible correspondences with NVSS:
the closest one, at a separation of 1\farcm29, is also the most interesting because it can 
be associated with the above X-ray source.
The \emph{WISE} counterpart has only $W1$ and $W2$ photometry, thus it is not possible to verify if
it lies within the strip.

\subsection{9Y-MST J1714$+$3227}   
There are 2 NVSS and 4 FIRST sources in the 6\arcmin\ region around this cluster.
The source closest to the centroid has very clear optical and mid-IR counterparts:
an elliptical galaxy without a particularly bright and blue nucleus, but with the
colours $W1 - W2 = 0.49$, $W2 - W3 = 1.65$ that place it at the lower end of the
\emph{WISE} Gamma-ray Strip, {and it is included in the KDEBLLACS catalogue.}

The other and brighter NVSS source does not have an optical counterpart, but there is
a \emph{WISE} source with a positional correspondence detected only in the $W1$ and $W2$
bandpasses. 
The former one is therefore the most likely candidate, 
but spectral data are required to confirm its blazar nature.

\subsection{9Y-MST J1744$+$4636}   
Three NVSS sources are within the angular distance of 6\arcmin. 
However, because it is a compact cluster with $R_\mathrm{max} = 3\farcm5$,  
only the closest source could be a reliable candidate.
It has a RASS and XMM counterpart, and the optical one in SDSS is a starlike object with
$u - r = 0.80$, but without spectral information.
The mid-IR counterpart has peculiar colours: $W1 - W2 = 0.45$, $W2 - W3 = 1.93$ locating
it in the BL Lac portion of the \emph{WISE} strip, whereas the very high $W3 - W4 = 3.74$ does
not confirm its position in the 3-colour strip. 
It is in the \cite{brescia15} list of candidate quasars.
The Spectral Energy Distribution appears dominated by the X-ray emission in the keV band,
around 10$^{-12}$ erg cm$^{-2}$~s$^{-1}$, higher than the optical emission by a factor of about 3,
typical of extreme HBL objects \citep{costamante01}.

\subsection{9Y-MST J2024$-$2234}  
This is a rather peculiar low-$g$ cluster likely corresponding to 4FGL J2025.3$-$2231
which has not any reported counterpart.
The analysis in a smaller region with $\Lambda_\mathrm{cut} = 0.7$ confirms the same structure,
while using $\Lambda_\mathrm{cut} = 0.5$ a cluster with 8 photons and $g = 2.908$ is found, 
as reported in Table~\ref{t:sources}.
The position of the latter cluster is much closer to two 4FGL sources and in the 6\arcmin\
radius there are two NVSS sources.
The faintest and more distant has a weak mid-IR possible counterpart and no more information
is available.
The closest has a possible optical-mid IR counterpart at about 4\arcsec, moreover it 
is in the low frequency flat spectrum radio sources LORCAT \citep{massaro14} catalogue.
The \emph{WISE} colours $W1 - W2 = 0.76$, $W2 - W3 = 1.98$, $W3 - W4 = 2.68$ locate it 
in the BL Lac region defined by \cite{dabrusco14}. 
Unfortunately no optical spectrum is available to support this classification.

\subsection{9Y-MST J2030$-$1622} 
This cluster corresponds to 4FGL J2030.9$-$1621 for which no counterpart is given.
There are three NVSS sources in its environment.
The brightest of them, at an angular distance from the centroid of 2\farcm5, does 
not have counterparts either in the optical nor in the mid-IR sky.
Therefore, no more indications on its nature can be inferred.
The closest source, at a distance of 49\arcsec, has a faint optical counterpart in the 
POSS corresponding to a mid-IR object with colours $W1 - W2 = 0.71$, 
$W2 - W3 = 1.91$, locating it in the BL Lac segment of the \emph{WISE} strip {and it is included in the KDEBLLACS sample.}

Finally, also the third and faintest NVSS source is without counterparts in the optical 
and mid-IR.
The second one is therefore currently the most interesting candidate, but spectral data 
will be useful for confirming its nature.

\subsection{9Y-MST 2046$-$5410} 
This cluster corresponds to 4FGL J2046.9$-$5409 source for which no counterpart was reported.
There are two SUMSS21 radio sources in its neighborhood at about the same angular distance:
the fainter source does not have a clear optical or mid-IR counterpart, while the brighter
one (angular distance of 3\farcm3) can be associated with a faint optical object also detected 
in the mid-IR.
Its \emph{WISE} colours are $W1 - W2 = 0.42$, $W2 - W3 = 2.09$ which are in the BL Lac
section of the \emph{WISE} strip.
Close to this position there is a PMN source, that, if associated with our candidate,
has a flux density that gives a marginally flat radio spectral index equal to $-0.55$.

\subsection{9Y-MST J2115$-$4938}  
This cluster has the highest $g$ in the sample and corresponds to 4FGL J2115.6$-$4938.
The proposed counterpart is MRSS 235-024179, a galaxy in the Muenster Red Sky Survey
including about 5.5 million galaxies with Galactic latitudes less than $-45 \degr$ \citep{ungruhe03}
There are two SUMSS21 sources within the search region: one of them is at a 
distance of 5\farcm5, well outside the $R_\mathrm{max} \approx 3\arcmin$ circle.
The closer source, at 1\farcm1, has a possible relatively bright optical counterpart 
well detected in the \emph{WISE} sky with the colours $W1 - W2 = 0.48$, $W2 - W3 = 2.10$ that place
it in the BL Lac segment of the strip.
Its position agree with that of the above galaxy.
Considering the radio flux density reported in PMN, it has a flat spectrum.
Thefore, it appears as a reliable blazar candidate.

\subsection{9Y-MST J2135$-$5759}   
There are three SUMSS21 sources near the centroid position of this object: the closest is at the
angular distance of only 0\farcm4, and another and brighter object, reported also in the 
CRATES catalogue, is at 2\farcm3.
Unfortunately, the latter source does not have a possible optical counterpart, and, 
taking into account the radio position uncertainty, there is a high source confusion 
in the \emph{WISE} mid-IR images.
Thus, the poor information on this object prevents any reliable identificaton, although 
it cannot be excluded as a possible counterpart of the high energy source.

The closer radio source has a possible optical and mid-IR counterpart at an angular 
distance of about 4\arcsec, that is higher than the SUMMSS21 position uncertainty of 
about 2\arcsec.
This object has the mid-IR colours $W1 - W2 = 0.40$, $W2 - W3 = 2.02$ placing it close
to end of the BL Lac segment of the $WISE$ strip.
The third SUMSS21 source, much fainter than the other two, is without possible
counterparts.
The occurrence of two possible counterparts requires a deeper investigation to disentangle
this ambiguity.

\subsection{9Y-MST J2240$-$1244} 
This cluster corresponds to 4FGL J2240.3$-$1246, that is associated with the RASS source
1RXS J224014.7-124736. 
Three NVSS sources are within the 6\arcmin\ search radius and all are at distances from the
centroid higher than 4\arcmin.
Only one has a possible optical counterpart and appears closely associated with the above 
X-ray source.
A local analysis with the much shorter $\Lambda_\mathrm{cut} = 0.3 \Lambda_m$ gives a more compact
cluster closer to this counterpart and marginally satisfying its $R_\mathrm{max}$ limit.
There is a mid-IR counterpart detected only in two bands and the resulting colour
$W1 - W2 = 0.43$ is compatible with a BL Lac object.

\subsection{9Y-MST J2240$-$4747} 
This cluster corresponds to 4FGL J2240.7$-$4746 and the reported `bcu' counterpart is the
radio source SUMSS J224042$-$474733.
This is only one SUMSS21 source in our search circle.
The possible optical counterpart has a distance offset of 3\farcs8, higher than the
nominal mean positional error of 1\farcs8; it has a bright \emph{WISE} counterpart detected
in all the four bands.
Its mid-IR colours are $W1 - W2 = 0.43$, $W2 - W3 = 2.04$, $W3 - W4 = 2.57$ well
compatible with the BL Lac segment of blazar strip.

\subsection{9Y-MST J2321$-$2606}  
Two NVSS sources are within the search radius: the brighter one is at a distance of 
1\farcm2 and the other at 4\farcm9.
The former source has a radio flux density brighter by about a factor of 6 and is
included in the HMQ catalogue \citep{flesch15} as a photometric quasar candidate with an
estimated $z$ of 0.7.
It is well detected in the mid-IR and the $WISE$ colours are $W1 - W2 = 0.73$, 
$W2 - W3 = 2.56$, locating the source near the centre of the WISE strip in the mixed
FSRQ and BL Lac objects region, {but it is included in the KDEBLLACS sample \citep{dabrusco19}.}

\section{Summary and discussion}\label{s:conclusions}

The extragalactic $\gamma$-ray sky appears dominated by blazar sources \citep{massaro16}
and therefore the discovery of new blazars can be driven by the search for 
counterparts of new high energy sources.
In the new $\gamma$-ray catalogues there are several sources not yet associated 
with extragalactic objects.
We extracted from the 9Y-MST catalogue a sample of 24 unassociated clusters for
which there is at least a radio source at an angular separation from the centroid 
lower than 6\arcmin.
For each of these sources we extended the search to other wavelength ranges to verify
if they exhibit some properties allowing a classification as blazar candidates.

Unfortunately, optical spectra were available for only three sources, while two \emph{WISE}
colours were obtained for 18 sources, which are located in the BL Lac portion of 
the blazar strip.
{We also verified if these possible counterparts are in the most recent catalogues
based on mid-IR data and found that only one is in the WIBRaLS2, while other 8 are
in the KDEBLLACS.
}
In any case, more observations to definitely confirm their nature are needed.

13 of these clusters were also found to be in the recent preliminary version of the
4FGL catalogue \citep{4FGL}, confirming the validity of MST findings.

{
It is interesting that about all objects are BL Lac candidates and the only one with a 
FSRQ bordeline position has a featureless SDSS spectrum.
This finding agrees with the results of a previous blazar search \citep{paperII} in which
only three objects in a sample of 30 exhibited mid-IR colours in the FSRQ region.
A possible explanation is that our cluster selection at energies higher than 10 GeV is biased
to extract BL Lacs, and particularly HBL sources, rather than FSRQs.
HBL objects, in fact, have the Compton component in their Spectral Energy Distribution peaking
in the GeV range, while for the other blazars it is at lower energies \citep{abdo10b,fan16}
and therefore faint hard sources are preferentially detected against a softer background.
Only the bright radio source PKS J0143$-$0119 (see Section 4.5), for which an optical spectrum is available,
exhibits some emission lines typical of FSRQs but its \emph{WISE} colours are in the BL Lac region,
indicating that it could be an outlier.
}
The main aim of the present work is to contribute to the knowledge of the BL Lac population, 
in particular for what concerns low brightness objects which are easily detected at high energies.
Following this approach, a possibility to be explored is to verify the existence of a
subclass of BL Lacs too faint in radio band \citep{massaro17}
to be marginally detected at the sensitivity level of the available surveys. 

\begin{sidewaystable*}
\caption{Coordinates and main properties of the MST clusters detected at energies higher 
than 10 GeV and of their possible candidate blazar counterparts.
Celestial coordinates are J2000, angular distances $\Delta\theta$ are 
computed between the centroids of MST clusters and those of indicated 
counterparts.
See the main text for discussion.}
\centering
{\small
\begin{tabular}{lrrrrrrrrrrl}
\hline
   9Y-MST     &   RA    &  Dec     & $n$ &  $g$  &  $M$  & $R_m$ & $R_\mathrm{max}$ &  RA  &   Dec      & $\Delta\theta$ &    Counterpart            \\
              &   deg   &  deg     &     &       &       &  $'$  &  $'$ &     $\degr$   &   $\degr$      & $'$     &                            \\
\hline
              &         &          &     &       &       &       &      &           &             & $'$     &                           \\
 J0013$-$3222 &   3.439 & $-$32.373 & 10 & 2.319 & 23.194 & 12.7 & 24.7 &   3.41413 & $-$32.41265 & 2.7 & ROXA J001339.1$-$322443.8     \\  
 J0024$+$2401 &   6.028 &    24.027 &  8 & 2.618 & 20.942 & 4.98 & 11.9 &           &             &     &                               \\  
              &   6.037 &    24.031 &  6 & 3.598 & 21.587 & 3.72 & 5.76 &   6.02542 &    24.07731 & 2.8 & SDSS J002406.10$+$240438.3  ($^a$)  \\
 J0122$+$1033 &  20.617 &    10.553 &  4 & 3.815 & 15.262 & 1.14 & 8.10 &  20.59842 &    10.53700 & 1.5 & SDSS J012223.62$+$103213.2    \\ 
 J0127$+$1737 &  21.984 &    17.627 &  4 & 6.428 & 25.710 & 1.38 & 3.30 &  22.01921 &    17.60536 & 2.3 & SDSS J012804.61$+$173619.3    \\ 
 J0143$-$0122 &  25.855 &  $-$1.373 &  5 & 3.541 & 17.706 & 2.52 & 6.30 &  25.81967 &  $-$1.31692 & 3.8 & SDSS J014316.72$-$011900.9    \\ 
 J0202$+$2942 &  30.637 &    29.709 &  9 & 2.953 & 26.581 & 7.62 & 16.0 &           &             &     &                               \\  
              &  30.633 &    29.722 &  4 & 10.26 & 41.052 & 0.42 & 2.58 &  30.66542 &    29.72361 & 1.8 & SDSS J020239.70$+$294325.0  ($^b$)  \\   
 J0332$+$8227 &  53.103 &    82.451 &  9 & 2.761 & 24.848 & 4.68 & 10.7 &  53.09908 &    82.44589 & 0.3 & AllWISE J033223.77$+$822645.1 \\ 
 J0557$+$7705 &  89.392 &    77.087 &  6 & 3.883 & 23.296 & 3.66 & 6.78 &  89.33942 &    77.07867 & 0.9 & AllWISE J055721.46$+$770443.2 \\ 
 J0650$-$5146 & 102.563 & $-$51.767 & 11 & 3.471 & 38.178 & 3.54 & 11.8 & 102.54304 & $-$51.73942 & 1.7 & AllWISE J065010.33$-$514421.9 \\ 
 J0752$+$7120 & 118.177 &    71.341 &  4 & 5.017 & 20.069 & 1.74 & 2.46 & 118.10550 &    71.34722 & 1.4 & AllWISE J075225.32$+$712050.0 \\ 
 J0947$+$1120 & 146.991 &    11.349 &  4 & 4.926 & 19.702 & 1.62 & 4.20 & 146.94129 &    11.33942 & 3.0 & SDSS J094745.91$+$112021.9    \\ 
 J1003$-$2139 & 150.894 & $-$21.661 &  5 & 5.611 & 28.057 & 1.98 & 5.64 & 150.92861 & $-$21.63592 & 2.4 & AllWISE J100342.86$-$213809.3 \\ 
 J1636$-$0456 & 249.104 &  $-$4.939 &  7 & 4.479 & 31.351 & 2.88 & 3.90 & 249.13367 &  $-$4.91833 & 2.1 & SDSS J163632.08$-$045506.0    \\ 
 J1646$-$0942 & 251.500 &  $-$9.708 & 10 & 3.583 & 35.831 & 3.54 & 7.62 & 251.50479 &  $-$9.68839 & 1.2 & AllWISE J164601.15$-$094118.2 \\ 
 J1714$+$3227 & 258.647 &    32.450 &  9 & 5.249 & 47.241 & 2.40 & 5.76 & 258.63783 &    32.46711 & 1.1 & SDSS J171433.08$+$322801.6    \\ 
 J1744$+$4636 & 266.048 &    46.603 &  4 & 4.560 & 18.239 & 1.80 & 3.54 & 266.01217 &    46.62797 & 2.1 & SDSS J174402.91$+$463740.7    \\ 
 J2024$-$2234 & 306.219 & $-$22.567 & 12 & 2.785 & 33.424 & 4.74 & 13.1 &           &             &     &                               \\ 
              & 306.301 & $-$22.515 &  8 & 2.908 & 23.262 & 3.48 & 9.36 & 306.31321 & $-$22.50511 & 1.0 & AllWISE J202515.17$-$223018.4  ($^c$) \\
 J2030$-$1622 & 307.714 & $-$16.377 &  9 & 3.014 & 27.127 & 2.76 & 9.18 & 307.70054 & $-$16.37331 & 1.1 & AllWISE J203048.13$-$162223.9 \\ 
 J2046$-$5410 & 311.686 & $-$54.175 &  9 & 4.166 & 37.496 & 2.22 & 8.64 & 311.75304 & $-$54.21264 & 3.3 & AllWISE J204700.73$-$541245.5 \\  
 J2115$-$4938 & 318.916 & $-$49.639 &  5 & 7.052 & 35.259 & 1.86 & 2.94 & 318.93613 & $-$49.65195 & 1.1 & AllWISE J211544.67$-$493907.0 \\ 
 J2135$-$5759 & 323.829 & $-$57.999 &  6 & 4.057 & 24.341 & 2.58 & 8.64 & 323.81706 & $-$57.99562 & 0.4 & AllWISE J213516.09$-$575944.2 \\ 
 J2240$-$1244 & 340.122 & $-$12.738 &  7 & 3.061 & 21.428 & 4.86 & 9.00 & 340.06300 & $-$12.79414 & 4.8 & AllWISE J224015.12$-$124738.9 \\  
              & 340.091 & $-$12.736 &  4 & 4.186 & 16.743 & 3.06 & 3.90 & 340.06300 & $-$12.79414 & 3.8 & AllWISE J224015.12$-$124738.9  ($^d$) \\  
 J2240$-$4747 & 340.185 & $-$47.785 &  8 & 3.858 & 30.863 & 1.80 & 17.3 & 340.17565 & $-$47.79177 & 0.6 & AllWISE J224042.15$-$474730.3 \\  
 J2321$-$2606 & 350.445 & $-$26.105 &  5 & 4.954 & 24.769 & 2.82 & 6.00 & 350.44357 & $-$26.08476 & 1.3 & AllWISE J232146.45$-$260505.1 \\ 
\hline
\end{tabular}

($^a$) Cluster parameters from a local analysis with $\Lambda_\mathrm{cut} = 0.4 \Lambda_m $;
($^b$) Unweighted coordinates and cluster parameters from a local analysis with $\Lambda_\mathrm{cut} = 0.6 \Lambda_m $; \,\,\,\,\,\,\,\,\,\,\,\,
($^c$) Cluster parameters from a local analysis with $\Lambda_\mathrm{cut} = 0.6 \Lambda_m $;  
($^d$) Cluster parameters from a local analysis with $\Lambda_\mathrm{cut} = 0.3 \Lambda_m $
}
\label{t:sources}
\end{sidewaystable*}

\begin{sidewaystable*}
\caption{Radio, mid-IR and optical photometric data of the blazar candidates in Table 1.
}
\centering
{\small
\begin{tabular}{lrrrrrrlr}
\hline
   Source                   &  $F_\mathrm{rad}$ &   $W1$   &   $W2$     &    $W3$    &   $W4$     &  $z$~~& ~~~$r$ &  $u$~~\\
                               &  mJy  &   mag      &   mag      &    mag     &    mag     &  mag  &  mag   &  mag  \\
\hline
                               &       &            &            &            &            &       &        &       \\ 
 ROXA J001339.1$-$322443.8     & 155.7 & 14.04 0.03 & 13.85 0.04 &            &            & 16.23 &        &       \\
 SDSS J002406.10$+$240438.3    & 11.3  & 15.44 0.05 & 14.49 0.06 & 11.91 0.25 &            & 19.41 & 20.05  & 21.05 \\
 SDSS J012223.62$+$103213.2    & 13.6  & 15.15 0.04 & 14.87 0.08 &            &            & 18.76 & 19.32  & 19.95 \\
 SDSS J012804.61$+$173619.3    &  7.1  & 15.15 0.04 & 14.37 0.05 & 12.02 0.27 &            & 20.16 & 20.77  & 21.41 \\
 SDSS J014316.72$-$011900.9    & 848.7 & 15.01 0.04 & 14.40 0.05 & 12.02 0.40 &            & 18.88 & 20.04  & 21.16 \\ 
 SDSS J020239.70$+$294325.0    & 10.1  & 14.16 0.03 & 13.42 0.03 & 10.97 0.12 &  8.93 0.45 & 17.84 & 18.36  & 19.22 \\ 
 AllWISE J033223.77$+$822645.1 &  6.6  & 14.62 0.03 & 13.81 0.03 & 12.16 0.29 &  9.02 0.45 &       & 19.26F &       \\ 
 AllWISE J055721.46$+$770443.2 & 33.3  & 14.16 0.03 & 13.20 0.03 & 10.94 0.11 &            &       & 19.2 G &       \\
 AllWISE J065010.33$-$514421.9 & 38s   & 13.97 0.02 & 13.37 0.03 & 11.25 0.10 &            &       & 17.8 F &       \\
 AllWISE J075225.32$+$712050.0 &  8.7  & 15.17 0.03 & 14.67 0.05 &            &            &       & 20.0 G &       \\
 SDSS J094745.91$+$112021.9    & 48.5  &            &            &            &            & 16.89 & 17.66  & 19.73 \\
 AllWISE J100342.86$-$213809.3 & 11.4  & 13.39 0.03 & 13.08 0.03 & 11.61 0.10 &            &       & 14.76F &       \\ 
 SDSS J163632.08$-$045506.0    & 50.7  & 14.06 0.03 & 13.49 0.03 & 11.43 0.19 &            & 17.63 & 18.74  & 21.15 \\
 AllWISE J164601.15$-$094118.2 & 16.4  & 13.80 0.03 & 13.35 0.03 &            &            &       & 18.76F &       \\
 SDSS J171433.08$+$322801.6    & 17.6  & 14.19 0.03 & 13.70 0.03 & 12.05 0.26 &            & 17.17 & 18.01  & 20.24 \\
 SDSS J174402.91$+$463740.7    &  2.8  & 15.37 0.03 & 14.93 0.05 & 13.00 ----- &  9.22 0.49 & 18.82 & 19.49  & 20.29 \\
 AllWISE J202515.17$-$223018.4 & 22.3  & 14.26 0.03 & 13.50 0.04 & 11.53 0.24 &  8.85 0.52 &       & 17.94F &       \\ 
 AllWISE J203048.13$-$162223.9 & 14.8  & 14.76 0.03 & 14.05 0.05 & 12.14 0.40 &            &       & 18.7 F &       \\
 AllWISE J204700.73$-$541245.5 & 99.5s & 14.69 0.03 & 14.27 0.06 & 12.18 0.35 &            &       & 19.52F &       \\  
 AllWISE J211544.67$-$493907.0 & 84.9s & 13.77 0.03 & 13.28 0.03 & 11.18 0.14 &            &       & 17.74F &       \\           
 AllWISE J213516.09$-$575944.2 & 65.2s & 14.81 0.04 & 14.46 0.06 & 12.70 0.43 &            &       & 18.47F &       \\
 AllWISE J224015.12$-$124738.9 & 19.7  & 14.02 0.03 & 13.59 0.04 &            &            &       & 16.88F &       \\  
 AllWISE J224042.15$-$474730.3 & 67.8s & 13.34 0.02 & 12.90 0.03 & 10.86 0.11 &  8.29 0.25 &       & 16.25F &       \\   
 AllWISE J232146.45$-$260505.1 & 24.3  & 14.57 0.03 & 13.83 0.04 & 11.28 0.20 &            &       & 19.36F &       \\ 
\hline
\end{tabular}

s: radio flux density from SUMSS21; ~~~ r: r mag from SDSS;  ~~~ F: F mag from GSC2.3; ~~~ G: G mag from STScI. \\
}
\label{t:photom}

\end{sidewaystable*}

\begin{acknowledgements}
We acknowledge use of archival Fermi data.
We made large use of the online version of the Roma-BZCAT and of
the scientific tools developed at the ASI Space Science Data Center (SSDC),
of the final release of 6dFGS archive, of the Sloan Digital Sky Survey
(SDSS) archive, of the NED database and other astronomical catalogues
distributed in digital form (Vizier and Simbad) at Centre de
Dates astronomiques de Strasbourg (CDS) at the Louis Pasteur University.
This publication makes also use of data products from the Wide-field 
Infrared Survey Explorer, which is a joint project of the University of California, 
Los Angeles, and the Jet Propulsion Laboratory/California Institute of Technology, 
funded by the National Aeronautics and Space Administration.
\end{acknowledgements}

\paragraph{Compliance with Ethical Standards} 
Conflict of Interest: The authors declare that they have no conflict of interest.
Ethical approval: This article does not contain any studies with human participants or animals performed by any of the authors.

\bibliographystyle{spr-mp-nameyear-cnd}
\bibliography{bibliography} 

\end{document}